\documentclass[aps,prd,reprint,showpacs,longbibliography,superscriptaddress,nofootinbib,titlepage]{revtex4-1} 

\usepackage{amsthm}
\usepackage{amssymb}   
\usepackage{mathtools} 
\usepackage{hyperref}
\hypersetup{colorlinks=true,linkcolor=blue,urlcolor=blue,citecolor=blue}
\usepackage{accents}
\usepackage{tensor}
\usepackage[cal=boondox]{mathalfa}
\usepackage{lipsum}

\usepackage[T1]{fontenc}

\begin{document}

\title{Polynomial $BF$-type action for general relativity and anti-self-dual gravity}

\author{Diego Gonzalez}
\email[]{diego.gonzalez@correo.nucleares.unam.mx}
\affiliation{Instituto de Ciencias Nucleares, Universidad Nacional Aut\'onoma de M\'exico, Apartado Postal 70-543, Ciudad de M\'exico, 04510, M\'exico}
\affiliation{Departamento de F\'{\i}sica, Cinvestav, Instituto Polit\'ecnico Nacional 2508, San Pedro Zacatenco,
	07360, Gustavo A. Madero, Ciudad de M\'exico, M\'exico}

\author{Mariano Celada}
\email[]{mcelada@fis.cinvestav.mx}
\affiliation{Departamento de F\'isica, Universidad Aut\'onoma Metropolitana Iztapalapa, San Rafael Atlixco 186, 09340 Ciudad de M\'exico, M\'exico}

\author{Merced Montesinos} 
\email[]{merced@fis.cinvestav.mx}
\affiliation{Departamento de F\'{\i}sica, Cinvestav, Instituto Polit\'ecnico Nacional 2508, San Pedro Zacatenco, 
	07360, Gustavo A. Madero, 
	Ciudad de M\'exico, M\'exico}


\date{\today}

\begin{abstract}
We report a gravitational $BF$-type action principle propagating two (complex) degrees of freedom that, besides the gauge connection and the $B$ field, only employs an additional Lagrange multiplier. The action depends on two parameters and remarkably is polynomial in the $B$ field. For a particular choice of the involved parameters the action provides an alternative description of (complex) general relativity with a nonvanishing cosmological constant, whereas another choice corresponds to anti-self-dual gravity. Generic values of the parameters produce ``close neighbors'' of general relativity, although there is a peculiar choice of the parameters that leads to a Hamiltonian theory with two scalar constraints. Given the nontrivial form of the resulting scalar constraint for these models, we consider a more general setting where the scalar constraint is replaced with an arbitrary analytic function of some fundamental variables and show that the Poisson algebra involving this constraint together with the Gauss and vector constraints of the Ashtekar formalism closes, thus generating an infinite family of gravitational models that propagate the same number of degrees of freedom as general relativity.
\end{abstract}


\maketitle

\section{Introduction} The pure connection formulation~\cite{kras2011106} expresses complex general relativity with a nonvanishing cosmological constant as a diffeomorphism-invariant gauge theory where the sole dynamical variable is an $SO(3,\mathbb{C})$-valued connection.~This action emerges after integrating out the auxiliary fields present in the Plebanski formulation~\cite{pleb1977118} (see Ref.~\cite{CGM_2015} for a rigorous derivation), and it is perhaps surprising that such an economical gauge formulation for general relativity exists. Since it involves the square root of a matrix quadratic in the curvature of the connection, its form is rather complicated, making it difficult to directly use this formulation in applications to quantum gravity.~However, several aspects of the pure connection formulation have been scouted out in a perturbative framework~\cite{Krasnprd84,DelfKrasn1,Delfino2015}.

An interesting feature of the pure connection formulation is that although we take as the starting point Plebanski's action, which is a $BF$ theory supplemented with constraints (see Ref.~\cite{cgmBFReview} for an overview of the relationship between $BF$ theories and general relativity), we end up obtaining an action principle that, apart from the dependence on a gauge connection, does not resemble the original one: all the $BF$ character of the initial action is completely destroyed during the process. In order to describe general relativity with few variables and still have a $BF$-type action after integrating out some fields in the action, a slight modification of Plebanski's action was proposed in Ref.~\cite{cgm-PlebLike-2016}, showing that it led to the action principle introduced in Ref.~\cite{Krasn2015} to describe general relativity as a $BF$ theory with a potential term depending only on the $B$ field. Nevertheless, the latter action also involves square roots of matrices, and so it seems that the appearance of these annoying roots is inexorable when we wish to describe general relativity with few variables.

In this paper we report a $BF$-type action embodying a class of gravitational models that propagate two (complex) physical degrees of freedom, the same number as general relativity before imposing the appropriate reality conditions~\cite{pleb1977118,Ashprl57.2244,AshPRD.36.1587} (see also Refs.~\cite{capo1991841,peld1994115}). As dynamical variables, the action only employs the $B$ field, an $SO(3,\mathbb{C})$ connection, and a scalar density playing the role of a Lagrange multiplier. The action itself does not require the presence of square roots of matrices nor the introduction of additional variables imposing extra constraints as in the case of the Plebanski formulation. Moreover, it depends on two free parameters and is constructed in such a way that the $B$ field enters in a polynomial fashion.~Notably, a particular choice of the involved parameters provides a polynomial action for complex general relativity with a nonvanishing cosmological constant. Likewise, another selection of these parameters also allows us to describe anti-self-dual gravity, and so the action reported in this paper, like the one posed in Ref.~\cite{cgm-PlebLike-2016}, provides a unified treatment of general relativity and anti-self-dual gravity. In fact, it is the parameter playing the role of the cosmological constant in the case of general relativity that causes the switch to anti-self-dual gravity when turned off. Later on, we perform the canonical analysis of the two-parameter action and show that in addition to the usual Gauss and vector constraints, we also have a nontrivial scalar constraint. We compute the constraint algebra and show that it closes, concluding that the theory indeed possesses two (complex) physical degrees of freedom per space point. In particular, in the cases of general relativity and anti-self-dual gravity we show that the corresponding scalar constraints reduce to those already found in the literature.~Furthermore, we find that another choice of the involved parameters, which seems not to play a significant role at the Lagrangian level, produces a Hamiltonian theory with two scalar constraints. Finally, motivated by the complicated form of the ensuing scalar constraint in the general case, we promote it to an arbitrary analytic function (with the right weight) depending on some fundamental quantities and show that even in that case the Poisson algebra among the modified scalar constraint and the Gauss and vector constraints closes, indicating that the associated theory still propagates the same number of degrees of freedom as before.

\section{Polynomial $BF$-type action}

Let $\mathcal{M}$ be an orientable four-dimensional manifold and consider an $SO(3,\mathbb{C})$ principal bundle over $\mathcal{M}$\footnote{$SO(3,\mathbb{C})$ is the structure group employed to describe the Lorentzian theory. In the Euclidean case, the structure group is just $SO(3)$ and all the quantities are real valued.}. We start by setting up the $BF$-type action principle 
\begin{eqnarray}
S[A,B,\eta]=\int_\mathcal{M} &&\left\{ B_i\wedge F^i[A] - \frac{b}{6} B_i\wedge B^i \right.  \nonumber\\
 && \hspace{3mm} +\underaccent{\tilde}{\eta} \left[{\rm Tr}\tilde{N}^2-\frac{a}{2} ({\rm Tr}\tilde{N})^2\right]  d^4x
   \biggr\}, \label{polyBF}
\end{eqnarray}
where $d^4x$ stands for the ordered product $dx^0\wedge dx^1\wedge dx^2\wedge dx^3$, $A^i$ is an $SO(3,\mathbb{C})$ connection with curvature $F^i[A]=d A^i + (1/2) {\varepsilon}^i{}_{jk}  A^j  \wedge A^k$, $B^i$ are three nondegenerate 2-forms in the sense that the symmetric matrix $\tilde{N}$ defined by $\tilde{N}^{ij}d^4x:=B^i\wedge B^j$ is nonsingular, $\underaccent{\tilde}{\eta}$ is a scalar density of weight $-1$ (the number of tildes ``$\sim$'' above or below a quantity specifies its weight) acting as a Lagrange multiplier, $a$ is a dimensionless parameter  and $b$ is a parameter with the same dimensions as the cosmological constant (in our approach the action as a whole has dimensions of the inverse of the cosmological constant). The internal indices $i,j,\ldots=1,2,3$ are raised and lowered with the three-dimensional Euclidean metric $\delta_{ij}$, and ${\varepsilon}_{ijk}$ is the Levi-Civita symbol (${\varepsilon}_{123}=+1$). It is worth pointing out that although the first line of Eq. \eqref{polyBF} is topological (it is just $BF$ theory with a volume term), the addition of the second line breaks part of this topological symmetry and endows the resulting theory with local dynamics, as we shall see below.

The equations of motion arising from the variation of Eq.~\eqref{polyBF} with respect to each independent variable are given~by
\begin{subequations}\label{eqmot}
	\begin{eqnarray}
	&\delta A&:\ DB^i:=dB^i+\tensor{\varepsilon}{^i_{jk}}A^j\wedge B^k=0,\label{DB}\\
	& \delta B&:\ F^i+4\underaccent{\tilde}{\eta} \tilde{N}^i{}_j B^j-2 a \underaccent{\tilde}{\eta} ({\rm Tr}\tilde{N}) B^i-\frac{b}{3} B^i=0,\label{eqB}\\
	&\delta\eta&: {\rm Tr}\tilde{N}^2-\frac{a}{2} ({\rm Tr}\tilde{N})^2=0\label{eqtr}.
	\end{eqnarray}
\end{subequations}
The first equation says that the 2-form $B^i$ is covariantly constant, a feature that is shared by all the formulations of general relativity in the framework of $BF$ theories~\cite{cgmBFReview}. The second equation expresses the curvature of the connection $A^i$ as a function of $B^i$. The third equation constitutes a constraint involving only the traces of the matrix~$\tilde{N}$. 

We assert that the action~(\ref{polyBF}) describes (complex) general relativity with a nonvanishing cosmological constant for $a=1$ and $b\neq 0$, whereas the case $a=1$ and $b=0$ leads to anti-self-dual gravity. For other values of the parameters $a$ and $b$ ($3a-2\neq0$), the action~(\ref{polyBF}) embodies a family of diffeomorphism-invariant theories that, as we shall see in the Hamiltonian approach, propagate two complex degrees of freedom. Surprisingly, the canonical analysis reveals that for $3a-2=0$ the theory possesses two scalar constraints. Let us see all this in detail.

\section{General relativity}\label{sectGR}

In this section we assume $a=1$ and $b\neq 0$. To establish that in this case Eq. (\ref{polyBF}) is an action for general relativity, it is enough to show that the ensuing equations of motion lead to Plebanski's equations for general relativity. After all, Plebanski's equations (plus some suitable reality conditions~\cite{pleb1977118}, which we must assume too) are equivalent to Einstein's equations with or without a cosmological constant where the Urbantke metric~\cite{urba198425} constructed from the $B$'s (which is conformally related to a Lorentzian metric) plays the role of the spacetime metric~\cite{capo1991841,cgmBFReview}. 

The first step consists in finding a set of 2-forms satisfying the simplicity constraint, which is the chief ingredient of Plebanski's approach. Let us introduce the following quantities:
\begin{subequations}
\begin{eqnarray}
&&\Sigma^i:=-\frac{6}{b} \underaccent{\tilde}{\eta}\left[ 2 \tilde{N}^i{}_jB^j - ({\rm Tr}\tilde{N}) B^i \right] ,\label{Pleb2form}\\
&&\Psi:=\frac{b^2}{36\underaccent{\tilde}{\eta} {\rm det}\tilde{N}}\left[\tilde{N}^2-\frac{1}{2} ({\rm Tr}\tilde{N}) \tilde{N}\right].\label{MatixPsi}
\end{eqnarray}
\end{subequations}
Making the wedge product of Eq.~(\ref{Pleb2form}) with itself and using the characteristic equation of the matrix $\tilde{N}$, namely
\begin{equation}
	\tilde{N}^3-({\rm Tr}\tilde{N})\tilde{N}^2+\frac{1}{2}\left[({\rm Tr}\tilde{N})^2-{\rm Tr}\tilde{N}^2\right]\tilde{N}-\det\tilde{N}=0, \label{charac}
\end{equation}
together with Eq.~(\ref{eqtr}), we obtain
\begin{eqnarray}
\Sigma^i\wedge\Sigma^j -  \left(\frac{12}{b}\underaccent{\tilde}{\eta}\right)^2 {\rm det}\tilde{N} d^4x\ \delta^{ij}=0. \label{Pleb1} 
\end{eqnarray}
Thus $\Sigma^i\wedge\Sigma^j\sim\delta^{ij}$, which is the simplicity constraint. Consequently, the 2-forms $\Sigma^i$ given by Eq.~(\ref{Pleb2form}) constitute the Plebanski 2-forms, that is, the ones satisfying the simplicity constraint. On the other hand, by appealing again to Eq. \eqref{charac}, the product of Eqs.~\eqref{Pleb2form} and \eqref{MatixPsi} yields
\begin{equation}
	\Psi^i{}_j\Sigma^j=\frac{b}{3}B^i.\label{PsiSigma}
\end{equation}
Then, by using this result and Eq.~\eqref{Pleb2form}, Eq.~\eqref{eqB} takes the form
\begin{eqnarray}
F^i=\left(\Psi^i{}_j - \frac{b}{3}\delta^i_j\right) \Sigma^j. \label{Pleb4} 
\end{eqnarray}
Taking the $SO(3,\mathbb{C})$ covariant derivative on both sides of Eq.~\eqref{Pleb4}, noting that Eqs. \eqref{PsiSigma} and \eqref{DB} imply $D(\Psi_{ij}\Sigma^j)=0$, and bearing in mind the Bianchi identity $DF^i=0$, we conclude that
\begin{eqnarray}
D\Sigma^i=0,\label{Pleb2} 
\end{eqnarray}
which means that $\Sigma^i$ is covariantly constant. Finally, notice that because of Eq.~\eqref{eqtr}, the matrix $\Psi$ is traceless:
\begin{eqnarray}
{\rm Tr}\Psi=0.\label{Pleb3} 
\end{eqnarray}
Equations~(\ref{Pleb1}), (\ref{Pleb4}), (\ref{Pleb2}), and (\ref{Pleb3}) are the Plebanski equations of motion for general relativity, which have been obtained from the set of equations \eqref{DB}-\eqref{eqtr} with $a=1$ and $b\neq 0$. In other words, for these values of the parameters $a$ and $b$, the action \eqref{polyBF} describes general relativity with a nonvanishing cosmological constant given by $\Lambda=-b$. On shell, $\Psi_{ij}$ gets identified with the self-dual part of the Weyl tensor. It is worth recalling that in order to make full contact with Einstein's equations we must impose the reality conditions on the $\Sigma$'s (the same as in Plebanski's case)  and introduce the Urbantke metric for them, which, because of Eq.~\eqref{PsiSigma}, turns out to be conformally related to the Urbantke metric defined by the $B$'s.

\section{Anti-self-dual gravity}\label{sectASDG}

We now focus our attention on the case $a=1$ and $b=0$. After solving Eq. (\ref{eqB}) for $F^i$, we compute the product $F^i\wedge F^j$, which, because of Eqs. \eqref{charac} and \eqref{eqtr}, yields
\begin{equation}
	F^i\wedge F^j -\frac{1}{3}F_k\wedge F^k\ \delta^{ij}=0,
\end{equation}
where $F_k\wedge F^k=48\underaccent{\tilde}{\eta}^2 {\rm det}\tilde{N} d^4x$. This is the so-called instanton equation~\cite{Capo-7-1-001,Torreprd41} and it is the defining feature of (conformally) anti-self-dual gravity. It says that the curvature of the gauge connection $A^i$ satisfies the simplicity constraint, thus making $F^i$ proportional to the Plebanski 2-forms $\Sigma^i$:
\begin{equation}
	F^i=\frac{\Lambda}{3}\Sigma^i,
\end{equation}
where $\Lambda$ is a nonvanishing constant that is identified with the cosmological constant. The consequence of this relation [cf.~Eq.~\eqref{Pleb4}] is that all the classical solutions of anti-self-dual gravity are solutions of Einstein's equations with a nonvanishing cosmological constant and vanishing self-dual Weyl curvature (they are known as gravitational instantons~\cite{Samuel1988}). In conclusion, for $a=1$ and $b=0$ the action (\ref{polyBF}) describes anti-self-dual gravity.

We remark that, according to the action (\ref{polyBF}) for $a=1$, the origin of the cosmological constant in general relativity is different from that in anti-self-dual gravity. For the former, the cosmological constant is identified with the negative of the nonvanishing parameter $b$ appearing in Eq. (\ref{polyBF}), while for the latter this parameter is set to zero and the cosmological constant is the proportionality factor relating $F^i$ and the Plebanski 2-forms. Therefore, the parameter $b$ switches between general relativity with a nonvanishing cosmological constant ($b\neq 0$) and anti-self-dual gravity ($b=0$). As far as we know, the case of general relativity with a vanishing cosmological constant is not encompassed by Eq. (\ref{polyBF}).

\section{Canonical analysis}\label{canaly}

For $a\neq 1$ the equations of motion \eqref{DB}-\eqref{eqtr} cannot be expressed in nice forms as in the previous cases, and hence the action (\ref{polyBF}) describes other models of gravity for $a\neq 1$ and arbitrary $b$. Since the same action describes general relativity with a nonvanishing cosmological constant for the particular choice $a=1$ and $b\neq0$, these models can be regarded as ``close neighbors'' of general relativity or deformations of it. (Even anti-self-dual gravity itself fits into this class.)

In order to dig deeper into the structure of the family of models depicted by Eq. (\ref{polyBF}), let us go to the canonical formalism. We assume that the spacetime manifold $\mathcal{M}$ has topology $\mathbb{R}\times\Omega$, where, for the sake of simplicity, $\Omega$ is a compact spatial 3-manifold without a boundary. Then, there exists a global time function $t$ such that the (spatial) 3-manifolds with constant $t$ have the topology of $\Omega$.~The coordinates adapted to this decomposition are chosen such that the spatial components of tensors are denoted by $a,b,\dots=1,2,3$, whereas the function $t$ labels their time components. By performing the $3+1$ decomposition of the action (\ref{polyBF}), we obtain
\begin{eqnarray}
&\hspace{-4mm}S[A,B,\eta]=&\int_{\mathbb{R}}dt\int_{\Omega}d^3x \biggl\{\tilde{\Pi}^{ai} \dot{A}_{ai} +A_{ti} \tilde{\mathcal{G}}^i \nonumber\\
&&+  B_{tai} \tilde{E}^{ai}  +\underaccent{\tilde}{\eta} \left[{\rm Tr}\tilde{N}^2-\frac{a}{2} ({\rm Tr}\tilde{N})^2\right]\biggr\},\label{Hamil1}
\end{eqnarray}
where we have defined $\tilde{\Pi}^{ai}:=(1/2)\tilde{\eta}^{abc} B_{bc}{}^i$, $\tilde{E}^{ai}=\tilde{B}^{ai}-(b/3)\tilde{\Pi}^{ai}$ with $\tilde{B}^{ai}:=(1/2)\tilde{\eta}^{abc}F^i{}_{bc}$, and $\tilde{\mathcal{G}}^i:=\mathcal{D}_a\tilde{\Pi}^{ai}$. Here, a dot over a variable indicates its time derivative and $\tilde{\eta}^{abc}$ ($\underaccent{\tilde}{\eta}_{abc}$) is the three-dimensional Levi-Civita symbol satisfying $\tilde{\eta}^{123}=1$~($\underaccent{\tilde}{\eta}_{123}=1$). 

Notice that the action \eqref{Hamil1} is a quadratic polynomial in the components of the $B$ field because
\begin{equation}
	\tilde{N}^{ij}=B_{ta}{}^i \tilde{\Pi}^{aj} + B_{ta}{}^j \tilde{\Pi}^{ai}.\label{eqNB}
\end{equation}
Furthermore, since no time derivatives of $B_{tai}$ appear in the action, we can simplify the analysis by getting rid of these variables through the use of their corresponding equation of motion. The variation of Eq.~(\ref{Hamil1}) with respect to $B_{tai}$ gives
\begin{eqnarray}
\tilde{E}^{ai}+4 \underaccent{\tilde}{\eta} \left[ \tilde{N}^i{}_j-\frac{a}{2} ({\rm Tr}\tilde{N}) \delta^i_j\right] \tilde{\Pi}^{aj} = 0. \label{Hamil2}
\end{eqnarray}
To continue, let us assume that $\tilde{\Pi}^{ai}$ is nonsingular; its inverse, denoted here by $\underaccent{\tilde}{\Pi}_{ai}$, fulfills $\underaccent{\tilde}{\Pi}_{ai}\tilde{\Pi}^{aj}=\delta^j_i$ and $\underaccent{\tilde}{\Pi}_{ai}\tilde{\Pi}^{bi}=\delta^b_a$.

The solution of Eq.~\eqref{Hamil2} involves the two cases discussed in what follows.

\subsection{Case $3a-2\neq0$}

Equation \eqref{Hamil2} is a linear system of nine equations for the nine unknowns $B_{tai}$. The system is notwithstanding degenerate and has rank six, implying that the solution for $B_{tai}$ can be written as
\begin{eqnarray}
&B_{tai}=&-\frac{1}{8 \underaccent{\tilde}{\eta}} \tilde{E}^{b}{}_{i} \underaccent{\tilde}{\Pi}_{b}{}^j \underaccent{\tilde}{\Pi}_{aj}+ \frac{a}{8(3a-2) \underaccent{\tilde}{\eta}} \tilde{E}^{bj} \underaccent{\tilde}{\Pi}_{bj} \underaccent{\tilde}{\Pi}_{ai} \nonumber\\
&& +\varepsilon_{ijk}\tilde{N}^k\underaccent{\tilde}{\Pi}_{a}{}^j, \label{HamilSol}
\end{eqnarray}
where $\tilde{N}^k$ is an arbitrary internal 3-vector (of weight +1). Accordingly, the first line of the previous relation corresponds to a particular solution of Eq. \eqref{Hamil2}, whereas the second line comprises the homogeneous solution. Plugging Eq. \eqref{HamilSol}  back into the action (\ref{Hamil1}), we obtain the (classically) equivalent action
\begin{eqnarray}
 S[A_{ai},&&\tilde{\Pi}^{ai},A_{ti},N^a,\underaccent{\tilde}{\eta}]=\int_{\mathbb{R}}dt\int_{\Omega}d^3x\biggl\{\tilde{\Pi}^{ai}\dot{A}_{ai}+A_{ti}\tilde{\mathcal{G}}^i\nonumber\\
&&+N^a\tilde{\mathcal{V}}_a - \frac{1}{16 \underaccent{\tilde}{\eta}} \left[ {\rm Tr}\psi^2 - \frac{a}{3a-2} ({\rm Tr}\psi)^2 \right] \biggr\},\label{Hamil3}
\end{eqnarray}
where we have defined $N^a:=-(\det\tilde{\Pi})^{-1}\tensor{\tilde{\Pi}}{^a_i}\tilde{N^i}$, $\mathcal{\tilde{V}}_a:=\tilde{\Pi}^{b}{}_i F^i{}_{ba}$, and $\psi^{ij}:=\tilde{E}^{ai} \underaccent{\tilde}{\Pi}_{a}{}^j$, the latter being a symmetric matrix as a consequence of Eq.~(\ref{Hamil2}). Using the characteristic equation for the matrix $\psi^{ij}$ [replacing $\tilde{N}$ by $\psi$ in Eq. \eqref{charac}] to rewrite the last term on the right, we can finally express the action (\ref{Hamil3}) as
\begin{eqnarray}
&& S[A_{ai},\tilde{\Pi}^{ai},A_{ti},N^a,\underaccent{\tilde}{M}]\nonumber\\
&&=\int_{\mathbb{R}}dt\int_{\Omega}d^3x\left(\tilde{\Pi}^{ai}\dot{A}_{ai}+A_{ti}\tilde{\mathcal{G}}^i+N^a\tilde{\mathcal{V}}_a+\underaccent{\tilde}{M}\tilde{\tilde{\mathcal{H}}}\right),\label{actham}
\end{eqnarray}
which exhibits that the canonical pair $(A_{ai},\tilde{\Pi}^{ai})$ parametrizes the phase space, while the variables $A_{ti}$, $N^a$ and $\underaccent{\tilde}{M}:=(3/8)( \underaccent{\tilde}{\eta} \det\tilde{\Pi})^{-1}$ appear linearly in the action and thus play the role of Lagrange multipliers imposing the constraints
\begin{subequations}\label{constraints}
	\begin{eqnarray}
	&\mathcal{\tilde{G}}^i=&\mathcal{D}_a\tilde{\Pi}^{ai}\approx 0,\label{gauss}\\
	&\mathcal{\tilde{V}}_a=&\tilde{\Pi}^{b}{}_i F^i{}_{ba}\approx 0,\label{Vect}\\
	&\tilde{\tilde{\mathcal{H}}}:=&\Pi BB -\frac{2}{3}b \Pi \Pi B + \frac{1}{9} b^2 \Pi \Pi \Pi \nonumber\\
	&& -\gamma (\Pi\Pi\Pi)^{-1} (\Pi\Pi B - \frac{1}{3} b \Pi\Pi\Pi)^2 \approx 0,\label{scalar}
	\end{eqnarray}
\end{subequations}
with $\gamma:=3(a-1)/(3a-2)$ and we have introduced the notation $\Pi\Pi B:=(1/6)\underaccent{\tilde}{\eta}_{abc}\varepsilon_{ijk}\tilde{\Pi}^{ai}\tilde{\Pi}^{bj}\tilde{B}^{ck}$, etc. Notice that the Gauss~$\mathcal{\tilde{G}}^i$  and vector $\mathcal{\tilde{V}}_a$ constraints have the same form as those of the Ashtekar formalism of general relativity. However, the scalar constraint $\tilde{\tilde{\mathcal{H}}}$ turns out to be more complicated than Ashtekar's one and depends on the parameters $a$ and $b$.~To better understand the set of constraints (\ref{gauss})-\eqref{scalar}, we shall split the family of theories described by Eq.~(\ref{polyBF}) into two sectors: one with $b\neq0$ containing general relativity and another with $b=0$ including anti-self-dual gravity. 

Let us first consider the sector with $b\neq0$. In this case we can perform the canonical transformation $(A_{ai},\tilde{\Pi}^{ai})\longmapsto(A_{ai},\tilde{\pi}^{ai}:=\tilde{\Pi}^{ai}-(3/b)\tilde{B}^{ai})$, under which the Gauss and vector constraints remain invariant, but the scalar constraint is promoted to
\begin{eqnarray}
&&\hspace{-5mm}\tilde{\tilde{\mathcal{H}}}=\frac{b}{3}   \left[  \pi \pi B +  \frac{b}{3} \pi \pi \pi \right.  \nonumber\\
&&\hspace{-5mm} \left. -\frac{\gamma}{3}  \frac{  ( 9\pi B B + 6b \pi\pi B + b^2 \pi \pi \pi)^2}{\left( 27 BBB + 27 b \pi BB + 9b^2 \pi \pi B + b^3 \pi \pi \pi  \right)} \right]  \approx 0.\label{HamilF1}
\end{eqnarray}
As expected, in the case of general relativity, which according to Sec. \ref{sectGR} corresponds to the choice $a=1$ and $b\neq 0$, we have $\gamma=0$, and Eq.~(\ref{HamilF1}) reduces to the usual scalar constraint of the Ashtekar formalism for general relativity, modulo the global factor $b/3$. This independently verifies that the action \eqref{polyBF} indeed describes general relativity for the above choice of the parameters. For generic values of $a$ (different from $a=2/3$), we are compelled to verify whether the Poisson algebra of the constraints closes or not, in which case new constraints could arise. Since the Gauss and vector constraints are not modified, the only nontrivial Poisson bracket we have to compute is that of the scalar constraint (\ref{HamilF1}) with itself. To that end, we introduce the smeared scalar constraint $H[{\underaccent{\tilde}{M}}]:=\int_{\Omega} d^3x \underaccent{\tilde}{M} \tilde{\tilde{\mathcal{H}}}$, where the test field $\underaccent{\tilde}{M}$ has weight $-1$. We must then compute the Poisson bracket 
\begin{eqnarray}
&&\left\{ H[{\underaccent{\tilde}{M}{}_1}],H[{\underaccent{\tilde}{M}{}_2}] \right\} \nonumber\\
&&= \int_{\Omega} d^3x \left( \frac{\delta H[{\underaccent{\tilde}{M}{}_1}]}{\delta A_{ai} }  \frac{\delta H[{\underaccent{\tilde}{M}{}_2}]}{\delta \tilde{\pi}^{ai} } - \frac{\delta H[{\underaccent{\tilde}{M}{}_2}]}{\delta A_{ai} }  \frac{\delta H[{\underaccent{\tilde}{M}{}_1}]}{\delta \tilde{\pi}^{ai} }\right), \label{poissonHamiltonian}
\end{eqnarray}
which has the same expression if the canonical variables $(A_{ai},\tilde{\Pi}^{ai})$ are used instead. After some algebra, we obtain
\begin{eqnarray}
\left\{ H[{\underaccent{\tilde}{M}{}_1}],H[{\underaccent{\tilde}{M}{}_2}] \right\} = \int_{\Omega} d^3x  \tilde{\tilde{Q}}^{ab}  \underaccent{\tilde}{\underaccent{\tilde}{M}}{}_b \mathcal{\tilde{V}}_a, \label{poissonb}
\end{eqnarray}
where $\underaccent{\tilde}{\underaccent{\tilde}{M}}{}_a:=\underaccent{\tilde}{M}{}_1\partial_a\underaccent{\tilde}{M}{}_2-\underaccent{\tilde}{M}{}_2\partial_a\underaccent{\tilde}{M}{}_1$ and $\tilde{\tilde{Q}}^{ab}$ is the symmetric tensor density given by
\begin{eqnarray}
&&\tilde{\tilde{Q}}^{ab}:=\left(\frac{b}{9}\right)^2 \left[ \tilde{\pi}^a{}_i \tilde{\pi}^{bi} \right. \nonumber\\
&&\left. - 3 \gamma  \alpha_0 \left( \alpha_1 \tilde{\pi}^a{}_i \tilde{\pi}^{bi} + \alpha_2 \tilde{\pi}^{(a|}{}_i \tilde{B}^{|b)i} + \alpha_3 \tilde{B}^a{}_i \tilde{B}^{bi} \right)\right], \label{Qab1}
\end{eqnarray}
with
\begin{subequations}
\begin{eqnarray}
&\alpha_0:=&\frac{ 9 \pi B B + 6b \pi\pi B + b^2 \pi \pi \pi}{\left( 27 BBB + 27b \pi BB + 9b^2 \pi \pi B + b^3 \pi \pi \pi  \right)^2}, \\
&\alpha_1:=&18b BBB + 3 (9-4\gamma) b^2 \pi BB \nonumber\\
&&+ 4 (3-2\gamma) b^3 \pi \pi B + \frac{1}{3}(5-4\gamma) b^4 \pi\pi\pi, \\
&\alpha_2:=&54 BBB + 36(3-2\gamma) b \pi BB \nonumber\\
&&+ 6 (9-8\gamma) b^2 \pi \pi B + 8(1-\gamma)b^3 \pi\pi\pi, \\
&\alpha_3:=&3 (3-4\gamma) \left( 9 \pi BB + 6b \pi \pi B + b^2  \pi\pi\pi\right).
\end{eqnarray}
\end{subequations}
Therefore, the constraint algebra closes and the constraints $\mathcal{\tilde{G}}^i$, $\mathcal{\tilde{V}}_a$, and $\tilde{\tilde{\mathcal{H}}}$ are first class. Thus, the kind of gravitational models with $b\neq0$ and $3a-2\neq0$ propagate two (complex) physical degrees of freedom. In particular, for general relativity we have $\gamma=0$, which implies that the Poisson bracket \eqref{poissonb} reduces to the usual one up to a global factor.~The quantity \eqref{Qab1} can actually be interpreted as a densitized version of the inverse of the spatial metric~\cite{Hojman,beng19913158,peld1994115}, and we see that its expression is rather nontrivial in general, although in the case of general relativity it takes the expected form. We point out that the algebra of constraints closes directly from Eqs.~(\ref{gauss})-\eqref{scalar} (see Sec.~\ref{arbitsc}); we only performed the previous canonical transformation in order to make the passing from the general Hamiltonian theory to the Ashtekar formalism of general relativity more straightforward.

On the other hand, for $b=0$ the scalar constraint~(\ref{scalar}) yields
\begin{equation}\label{scASDG}
	\tilde{\tilde{\mathcal{H}}}=\Pi BB -\gamma (\Pi\Pi\Pi)^{-1} (\Pi\Pi B)^2\approx0,
\end{equation}
and it is readily seen that in the case of anti-self-dual gravity, namely $a=1$ according to Sec. \ref{sectASDG} (equivalently $\gamma=0$), we recover the expression for the scalar constraint found in Ref.~\cite{cgm-PlebLike-2016}. The Poisson bracket of Eq.~\eqref{scASDG} with itself then takes the same form as Eq.~\eqref{poissonb}, but with $\tilde{\tilde{Q}}^{ab}$ this time given by
\begin{eqnarray}
 \tilde{\tilde{Q}}^{ab}=\frac{1}{9}\tilde{B}^a{}_i \tilde{B}^{bi}-\frac{\gamma}{9}\frac{\Pi\Pi B}{(\Pi\Pi\Pi)^2}&&\left[(3-4\gamma)(\Pi\Pi B)\ \tilde{\Pi}^a{}_i \tilde{\Pi}^{bi}\right.\nonumber\\
&&+2(\Pi\Pi\Pi)\ \tilde{\Pi}^{(a|}{}_i \tilde{B}^{|b)i}\Bigr].\label{Qab2}
\end{eqnarray}
Thus, the Poisson algebra among the constraints $\mathcal{\tilde{G}}^i$, $\mathcal{\tilde{V}}_a$, and $\tilde{\tilde{\mathcal{H}}}$ closes, and so they are first class. Because of this, these models also propagate two (complex) physical degrees of freedom. Notice that in the case of anti-self-dual gravity (the inverse of) the spatial metric is constructed solely from the curvature, $\tilde{\tilde{Q}}^{ab}|_{\text{ASDG}}=(1/9)\tilde{B}^a{}_i \tilde{B}^{bi}$, whereas for general relativity its simpler form is quadratic in the canonical variable $\tilde{\Pi}^a{}_i$.

\subsection{Case $3a-2=0$}
This case is quite special, since from the Lagrangian point of view, nothing interesting seems to happen in the action \eqref{polyBF} at the particular value $a=2/3$. Actually, as far as the equations of motion are concerned, we have not found a way to express them in a closed form as we did for the case of general relativity.

For $a=2/3$, we find, from Eq. \eqref{Hamil2}, that the matrix $\psi_{ij}=\tilde{E}^a{}_i\underaccent{\tilde}{\Pi}_{aj}$ is traceless symmetric (it is proportional to the trace-free part of $\tilde{N}_{ij}$). Since this matrix is constructed out of the phase-space variables, this means that there is an additional constraint coming from the traceless property; the matrix $\psi_{ij}$ being symmetric is just another way of establishing the vector constraint. In this case, the action \eqref{Hamil1} can be written as
\begin{eqnarray}
	S[A_{ai},&&\tilde{\Pi}^{ai},A_{ti},\underaccent{\tilde}{\eta},N^a,\tilde{\rho}]=\int_{\mathbb{R}}dt\int_{\Omega}d^3x\biggl(\tilde{\Pi}^{ai}\dot{A}_{ai}\nonumber\\
	&&\left.+A_{ti}\tilde{\mathcal{G}}^i+N^a\tilde{\mathcal{V}}_a+\tilde{\rho}{\rm Tr}\psi-\frac{1}{16\underaccent{\tilde}{\eta}}{\rm Tr}\psi^2\right),\label{action2/3}
\end{eqnarray}
where $N^a$ and $\tilde{\rho}$ are Lagrange multipliers imposing the constraints mentioned above. In turn, the variable $\underaccent{\tilde}{\eta}$ imposes the constraint ${\rm Tr}\psi^2$, which, together with the constraint ${\rm Tr}\psi=0$ and the characteristic equation for $\psi_{ij}$, implies that ${\rm Tr}\psi^{-1}=0$.~(Alternatively, we can redefine the variable $\tilde{\rho}$ so that ${\rm Tr}\psi=0$ and ${\rm Tr}\psi^{-1}=0$ are the resulting constraints.)~Using the definition of $\psi_{ij}$, the action \eqref{action2/3} acquires the form
\begin{eqnarray}
S[A_{ai},&&\tilde{\Pi}^{ai},A_{ti},N^a,\underaccent{\tilde}{M}{}_1,\underaccent{\tilde}{M}{}_2]=\int_{\mathbb{R}}dt\int_{\Omega}d^3x\Bigl(\tilde{\Pi}^{ai}\dot{A}_{ai}\nonumber\\
&&\left.+A_{ti}\tilde{\mathcal{G}}^i+N^a\tilde{\mathcal{V}}_a+\underaccent{\tilde}{M}{}_1\tilde{\tilde{\mathcal{H}}}_1+\underaccent{\tilde}{M}{}_2\tilde{\tilde{\mathcal{H}}}_2\right),\label{action2/3_1}
\end{eqnarray}
where $\underaccent{\tilde}{M}{}_1:=(3/8)( \underaccent{\tilde}{\eta} \Pi\Pi\Pi)^{-1}$ and $\underaccent{\tilde}{M}{}_2:=3\tilde{\rho}(\Pi\Pi\Pi)^{-1}$, while $\tilde{\tilde{\mathcal{H}}}_1$ and $\tilde{\tilde{\mathcal{H}}}_2$ are respectively given by
\begin{subequations}
\begin{eqnarray}
	&&\tilde{\tilde{\mathcal{H}}}_1:=\Pi BB-\frac{2}{3}b \Pi \Pi B + \frac{1}{9} b^2 \Pi \Pi \Pi\approx 0,\label{scal1}\\
	&&\tilde{\tilde{\mathcal{H}}}_2:=\Pi \Pi B -  \frac{b}{3} \Pi \Pi \Pi\approx 0.\label{scal2}
\end{eqnarray}
\end{subequations}
Notice that $\tilde{\tilde{\mathcal{H}}}_1$ corresponds to Eq. \eqref{scalar} with $\gamma=0$ (or $a=1$, which is the case of general relativity if $b\neq0$), whereas $\tilde{\tilde{\mathcal{H}}}_2$ is the scalar constraint of the Ashtekar formalism for general relativity. Thus, the action principle \eqref{polyBF} for $a=2/3$ and $b\neq0$ somehow manages to include at the same time both scalar constraints for general relativity with a nonvanishing cosmological constant. It is worth realizing that for $b\neq 0$ the canonical transformation $(A_{ai},\tilde{\Pi}^{ai})\longmapsto(A_{ai},\tilde{\pi}^{ai}=\tilde{\Pi}^{ai}-(3/b)\tilde{B}^{ai})$ allows us to relate both constraints:
\begin{subequations}
\begin{eqnarray}
	&&\tilde{\tilde{\mathcal{H}}}_1(A_{ai},\tilde{\Pi}^{ai};b)=\frac{b}{3}\tilde{\tilde{\mathcal{H}}}_2(A_{ai},\tilde{\pi}^{ai};-b),\\
	&&\tilde{\tilde{\mathcal{H}}}_2(A_{ai},\tilde{\Pi}^{ai};b)=-\frac{3}{b}\tilde{\tilde{\mathcal{H}}}_1(A_{ai},\tilde{\pi}^{ai};-b).
\end{eqnarray}
\end{subequations}
On the other hand, for $b=0$ $\tilde{\tilde{\mathcal{H}}}_1$ becomes the scalar constraint for anti-self-dual-gravity, while $\tilde{\tilde{\mathcal{H}}}_2$ reduces to the Ashtekar scalar constraint for general relativity with a vanishing cosmological constant (there is no way to relate these constraints by using the above canonical transformation). Whether the simultaneous presence of both constraints leads to a consistent Hamiltonian theory for the different choices of $b$ is still being explored.

\section{Models with an arbitrary scalar constraint}\label{arbitsc}
As we have seen in Sec.~\ref{canaly}, the scalar constraint corresponding to the action principle \eqref{polyBF}, namely Eq. \eqref{scalar} or Eq. \eqref{HamilF1} for $b\neq 0$, is a rather complicated function of the fundamental blocks $\Pi\Pi\Pi$, $\Pi\Pi B$, $\Pi BB$, and $BBB$. Nevertheless, the set made up of this constraint together with the Gauss and vector constraints, Eqs. \eqref{gauss} and \eqref{Vect} respectively, is closed under the Poisson bracket. As a function of the previous fundamental blocks, how general can the scalar constraint be in order to form a closed set with the Gauss and vector constraints?~In this section we show that we can consider as a scalar constraint any analytic function (of weight $+2$) of the aforementioned fundamental blocks and still have a closed constraint algebra. Notice that particular instances of this statement have already been established in the literature~\cite{capo198963} (see also Refs.~\cite{beng1991254,KransDef}).

To proceed, let us consider a scalar constraint of the form
\begin{equation}\label{newscal}
	\tilde{\tilde{\mathcal{H}}}=\tilde{\tilde{\mathcal{H}}}(\Pi\Pi\Pi,\Pi\Pi B,\Pi BB,BBB)\approx 0,
\end{equation}
which is analytic in the four arguments. Since the Gauss and vector constraints form a closed set under the Poisson bracket by themselves, we just have to compute the Poisson bracket involving them and the scalar constraint \eqref{newscal} and, of course, the bracket of Eq.~\eqref{newscal} with itself.~We now introduce the smeared versions of the Gauss and vector constraints as $G[\Lambda]:=\int_{\Omega} d^3 x \Lambda_i\tilde{\mathcal{G}}^i$ and $V[N]:=\int_{\Omega} d^3 x N^a\tilde{\mathcal{V}}_a$, respectively, where the internal vector $\Lambda_i$ and the spatial tangent vector $N^a$ play the role of test functions. Note that fundamental blocks constructed from $\tilde{\Pi}^{ai}$ and $\tilde{B}^{ai}$ are internal scalars, and so the scalar constraint \eqref{newscal} is also an internal scalar. Hence, it Poisson commutes with the Gauss constraint. 

For any functional $F[A,\tilde{\Pi}]$ of the phase-space variables $(A_{ai},\tilde{\Pi}^{ai})$, the action of the vector constraint on it is given by
\begin{equation}
	\{V[N],F[A,\tilde{\Pi}]\}=\pounds_NF[A,\tilde{\Pi}]-\{G[\rho],F[A,\tilde{\Pi}]\},\label{vectF}
\end{equation}
where $\rho_i:=N^aA_{ai}$ and $\pounds_N$ stands for the Lie derivative along the vector field $N^a$, which in this case is understood as the result of a functional variation, that is,
\begin{equation}
	\pounds_NF=\int_{\Omega}d^3x\left(\pounds_NA_{ai}\frac{\delta }{\delta A_{ai}}+\pounds_N\tilde{\Pi}^{ai}\frac{\delta }{\delta \tilde{\Pi}^{ai}}\right)F.
\end{equation}
To get rid of the second term on the right-hand side of Eq. \eqref{vectF} it is customary to supersede the vector constraint by the diffeomorphism constraint $\tilde{\mathcal{D}}_a:=\tilde{\mathcal{V}}_a+A_{ai}\tilde{\mathcal{G}}^i$. The latter generates spatial diffeomorphisms according to the rule
\begin{equation}
\{D[N],F[A,\tilde{\Pi}]\}=\pounds_NF[A,\tilde{\Pi}],
\end{equation}
where $D[N]:=\int_{\Omega} d^3 x N^a\tilde{\mathcal{D}}_a$.

Using Eq. \eqref{vectF}, the Poisson bracket between the vector constraint and the constraint \eqref{newscal} can be written as
\begin{equation}
	\hspace{-2mm}\{V[N],H[\underaccent{\tilde}{M}]\}=H[-\pounds_N\underaccent{\tilde}{M}]+G[\theta]
\end{equation}
for $\theta_i:=-\underaccent{\tilde}{M}N^a(\partial \tilde{\tilde{\mathcal{H}}}/\partial\tilde{\Pi}^a{}_i)$, and thereby it closes. Finally, some algebra shows that the Poisson bracket between Eq. \eqref{newscal} and itself takes the same form as Eq. \eqref{poissonb} with the following expression for the spatial metric $\tilde{\tilde{Q}}^{ab}$:
\begin{eqnarray}
	&&\tilde{\tilde{Q}}^{ab}=\frac{1}{3}\left[-\frac{\partial\tilde{\tilde{\mathcal{H}}}}{\partial\Pi\Pi\Pi}\frac{\partial\tilde{\tilde{\mathcal{H}}}}{\partial\Pi BB}+\frac{1}{3}\left(\frac{\partial\tilde{\tilde{\mathcal{H}}}}{\partial\Pi\Pi B}\right)^2\right]\tilde{\Pi}^a{}_i \tilde{\Pi}^{bi}\nonumber\\
	&&+\left[-\frac{\partial\tilde{\tilde{\mathcal{H}}}}{\partial\Pi\Pi\Pi}\frac{\partial\tilde{\tilde{\mathcal{H}}}}{\partial BBB}+\frac{1}{9}\frac{\partial\tilde{\tilde{\mathcal{H}}}}{\partial\Pi\Pi B}\frac{\partial\tilde{\tilde{\mathcal{H}}}}{\partial \Pi BB}\right]\tilde{\Pi}^{(a|}{}_i \tilde{B}^{|b)i}\nonumber\\
	&&+\frac{1}{3}\left[-\frac{\partial\tilde{\tilde{\mathcal{H}}}}{\partial\Pi\Pi B}\frac{\partial\tilde{\tilde{\mathcal{H}}}}{\partial BBB}+\frac{1}{3}\left(\frac{\partial\tilde{\tilde{\mathcal{H}}}}{\partial\Pi BB}\right)^2\right]\tilde{B}^a{}_i \tilde{B}^{bi}.\label{genQ}
\end{eqnarray}
Accordingly, the Poisson algebra generated by the Gauss constraint \eqref{gauss}, the vector constraint \eqref{Vect}, and the general scalar constraint \eqref{newscal} closes. In consequence, a theory subject only to this set of first-class constraints still propagates the same number of degrees of freedom as general relativity, thus giving rise to a huge family of gravitational models propagating two degrees of freedom, one for each noncanonically equivalent choice of the scalar constraint \eqref{newscal}.~One of such models is the one embodied in the scalar constraint \eqref{scalar}, which in turn produces different gravitational models (including general relativity) depending on the value of the parameters contained in the original polynomial action \eqref{polyBF}. To check this, we can easily verify that the spatial metric \eqref{genQ} yields Eqs.~\eqref{Qab1} and \eqref{Qab2} for their corresponding scalar constraints \eqref{HamilF1} and \eqref{scASDG}.

\section{Conclusions}

In this paper we have posed a new $BF$-type action for general relativity with a nonvanishing cosmological constant that is polynomial in the $B$ field. The action itself turns out to be a particular member of a family of gravitational models depicted by the action \eqref{polyBF}, which depends on two parameters $a$ and $b$. According to Sec. \ref{sectGR}, general relativity with a nonvanishing cosmological constant corresponds to the choice $a=1$ and $b\neq 0$, where (the negative of) the latter gets identified with the cosmological constant. Likewise, in Sec. \ref{sectASDG} we showed that for $a=1$ and $b=0$ the action \eqref{polyBF} describes anti-self-dual gravity. Thus, the parameter $b$ switches between general relativity and anti-self-dual gravity when $a=1$. As far as we know, general relativity with a vanishing cosmological constant cannot be described using the action \eqref{polyBF}, although it would be nice to have an analogous polynomial description of it to complete the landscape.

We also performed the canonical analysis of the $BF$-type action in Sec. \ref{canaly} and established that for $3a-2\neq0$ the theory propagates two (complex) physical degrees of freedom, which applies to general relativity and anti-self-dual-gravity as well. We point out that this restraint on the value of $a$ only shows up at the Hamiltonian level, whereas at the Lagrangian level taking $a=2/3$ in the action \eqref{polyBF} does not seem to create any obstacles (although no closed form for the equations of motion was found). The case $a=2/3$ is actually something special, since the resulting canonical theory has two scalar constraints, namely Eqs. \eqref{scal1} and \eqref{scal2}. The consistency of the presence of these two constraints and their meaning is currently being investigated. It is worth mentioning that from the Lagrangian standpoint, this model can also be regarded as Plebanski's action $S_{\text{Pl}} [A,B,\Psi,\rho]$ plus an additional term imposing either the constraint ${\rm Tr}\Psi^2=0$ or ${\rm Tr}\Psi^{-1}=0$.

In addition, based on the nontrivial form of the scalar constraint resulting from the canonical analysis of Sec. \ref{canaly}, we generalized it by allowing as a scalar constraint an arbitrary analytic function of the fundamental blocks $\Pi\Pi\Pi$, $\Pi\Pi B$, $\Pi BB$, and $BBB$, and showed in Sec.~\ref{arbitsc} that even in that case the Gauss, vector, and new scalar constraints form a closed set under the Poisson bracket. The constraints are then first class and the theory propagates the same number of degrees of freedom as before. Thus, we have an infinite family of gravitational models propagating the same number of degrees of freedom as general relativity, some of which have already been explored in the literature and have a Lagrangian counterpart~\cite{capo198963,beng1991254,KransDef,PhysRevD.84.024034}.

Looking ahead into the future, we think the action \eqref{polyBF}, being polynomial in the $B$ field, might become a good candidate to explore a nonperturbative path integral quantization of gravity. The fact that Eq.~\eqref{polyBF} actually gives a family of ``close neighbors'' to general relativity could be really helpful to attain this goal, since we can select among them one or several gravitational models with nice theoretical features such as renormalizability that might render them more amenable to quantization.\\

\acknowledgements

This work was partially supported by Consejo Nacional de Ciencia y Tecnolog\'{i}a (CONACyT), M\'{e}xico, Grants Nos. 237004-F and 237351. M.C. would like to thank the financial support of PRODEP Grant No. 12313153 (through UAM-I). D.G. is supported with a DGAPA-UNAM postdoctoral fellowship.

\bibliography{references}

\begin{thebibliography}{23}%
\makeatletter
\providecommand \@ifxundefined [1]{%
 \@ifx{#1\undefined}
}%
\providecommand \@ifnum [1]{%
 \ifnum #1\expandafter \@firstoftwo
 \else \expandafter \@secondoftwo
 \fi
}%
\providecommand \@ifx [1]{%
 \ifx #1\expandafter \@firstoftwo
 \else \expandafter \@secondoftwo
 \fi
}%
\providecommand \natexlab [1]{#1}%
\providecommand \enquote  [1]{``#1''}%
\providecommand \bibnamefont  [1]{#1}%
\providecommand \bibfnamefont [1]{#1}%
\providecommand \citenamefont [1]{#1}%
\providecommand \href@noop [0]{\@secondoftwo}%
\providecommand \href [0]{\begingroup \@sanitize@url \@href}%
\providecommand \@href[1]{\@@startlink{#1}\@@href}%
\providecommand \@@href[1]{\endgroup#1\@@endlink}%
\providecommand \@sanitize@url [0]{\catcode `\\12\catcode `\$12\catcode
  `\&12\catcode `\#12\catcode `\^12\catcode `\_12\catcode `\%12\relax}%
\providecommand \@@startlink[1]{}%
\providecommand \@@endlink[0]{}%
\providecommand \url  [0]{\begingroup\@sanitize@url \@url }%
\providecommand \@url [1]{\endgroup\@href {#1}{\urlprefix }}%
\providecommand \urlprefix  [0]{URL }%
\providecommand \Eprint [0]{\href }%
\providecommand \doibase [0]{http://dx.doi.org/}%
\providecommand \selectlanguage [0]{\@gobble}%
\providecommand \bibinfo  [0]{\@secondoftwo}%
\providecommand \bibfield  [0]{\@secondoftwo}%
\providecommand \translation [1]{[#1]}%
\providecommand \BibitemOpen [0]{}%
\providecommand \bibitemStop [0]{}%
\providecommand \bibitemNoStop [0]{.\EOS\space}%
\providecommand \EOS [0]{\spacefactor3000\relax}%
\providecommand \BibitemShut  [1]{\csname bibitem#1\endcsname}%
\let\auto@bib@innerbib\@empty
\bibitem [{\citenamefont {Krasnov}(2011{\natexlab{a}})}]{kras2011106}%
  \BibitemOpen
  \bibfield  {author} {\bibinfo {author} {\bibfnamefont {K.}~\bibnamefont
  {Krasnov}},\ }\href {\doibase 10.1103/PhysRevLett.106.251103} {\bibfield
  {journal} {\bibinfo  {journal} {Phys. Rev. Lett.}\ }\textbf {\bibinfo
  {volume} {106}},\ \bibinfo {pages} {251103} (\bibinfo {year}
  {2011}{\natexlab{a}})}\BibitemShut {NoStop}%
\bibitem [{\citenamefont {Pleba\'{n}ski}(1977)}]{pleb1977118}%
  \BibitemOpen
  \bibfield  {author} {\bibinfo {author} {\bibfnamefont {J.~F.}\ \bibnamefont
  {Pleba\'{n}ski}},\ }\href {\doibase 10.1063/1.523215} {\bibfield  {journal}
  {\bibinfo  {journal} {J. Math. Phys.}\ }\textbf {\bibinfo {volume} {18}},\
  \bibinfo {pages} {2511} (\bibinfo {year} {1977})}\BibitemShut {NoStop}%
\bibitem [{\citenamefont {Celada}\ \emph {et~al.}(2015)\citenamefont {Celada},
  \citenamefont {Gonz\'alez},\ and\ \citenamefont {Montesinos}}]{CGM_2015}%
  \BibitemOpen
  \bibfield  {author} {\bibinfo {author} {\bibfnamefont {M.}~\bibnamefont
  {Celada}}, \bibinfo {author} {\bibfnamefont {D.}~\bibnamefont {Gonz\'alez}},
  \ and\ \bibinfo {author} {\bibfnamefont {M.}~\bibnamefont {Montesinos}},\
  }\href {\doibase 10.1103/PhysRevD.92.044059} {\bibfield  {journal} {\bibinfo
  {journal} {Phys. Rev. D}\ }\textbf {\bibinfo {volume} {92}},\ \bibinfo
  {pages} {044059} (\bibinfo {year} {2015})}\BibitemShut {NoStop}%
\bibitem [{\citenamefont {Krasnov}(2011{\natexlab{b}})}]{Krasnprd84}%
  \BibitemOpen
  \bibfield  {author} {\bibinfo {author} {\bibfnamefont {K.}~\bibnamefont
  {Krasnov}},\ }\href {\doibase 10.1103/PhysRevD.84.024034} {\bibfield
  {journal} {\bibinfo  {journal} {Phys. Rev. D}\ }\textbf {\bibinfo {volume}
  {84}},\ \bibinfo {pages} {024034} (\bibinfo {year}
  {2011}{\natexlab{b}})}\BibitemShut {NoStop}%
\bibitem [{\citenamefont {Delfino}\ \emph
  {et~al.}(2015{\natexlab{a}})\citenamefont {Delfino}, \citenamefont
  {Krasnov},\ and\ \citenamefont {Scarinci}}]{DelfKrasn1}%
  \BibitemOpen
  \bibfield  {author} {\bibinfo {author} {\bibfnamefont {G.}~\bibnamefont
  {Delfino}}, \bibinfo {author} {\bibfnamefont {K.}~\bibnamefont {Krasnov}}, \
  and\ \bibinfo {author} {\bibfnamefont {C.}~\bibnamefont {Scarinci}},\ }\href
  {\doibase 10.1007/JHEP03(2015)118} {\bibfield  {journal} {\bibinfo  {journal}
  {J. High Energy Phys.}\ }\textbf {\bibinfo {volume} {03}},\ \bibinfo {pages}
  {118} (\bibinfo {year} {2015}{\natexlab{a}})}\BibitemShut {NoStop}%
\bibitem [{\citenamefont {Delfino}\ \emph
  {et~al.}(2015{\natexlab{b}})\citenamefont {Delfino}, \citenamefont
  {Krasnov},\ and\ \citenamefont {Scarinci}}]{Delfino2015}%
  \BibitemOpen
  \bibfield  {author} {\bibinfo {author} {\bibfnamefont {G.}~\bibnamefont
  {Delfino}}, \bibinfo {author} {\bibfnamefont {K.}~\bibnamefont {Krasnov}}, \
  and\ \bibinfo {author} {\bibfnamefont {C.}~\bibnamefont {Scarinci}},\ }\href
  {\doibase 10.1007/JHEP03(2015)119} {\bibfield  {journal} {\bibinfo  {journal}
  {J. High Energy Phys.}\ }\textbf {\bibinfo {volume} {03}},\ \bibinfo {pages}
  {119} (\bibinfo {year} {2015}{\natexlab{b}})}\BibitemShut {NoStop}%
\bibitem [{\citenamefont {Celada}\ \emph
  {et~al.}(2016{\natexlab{a}})\citenamefont {Celada}, \citenamefont
  {Gonz\'alez},\ and\ \citenamefont {Montesinos}}]{cgmBFReview}%
  \BibitemOpen
  \bibfield  {author} {\bibinfo {author} {\bibfnamefont {M.}~\bibnamefont
  {Celada}}, \bibinfo {author} {\bibfnamefont {D.}~\bibnamefont {Gonz\'alez}},
  \ and\ \bibinfo {author} {\bibfnamefont {M.}~\bibnamefont {Montesinos}},\
  }\href {http://stacks.iop.org/0264-9381/33/i=21/a=213001} {\bibfield
  {journal} {\bibinfo  {journal} {Classical Quantum Gravity}\ }\textbf
  {\bibinfo {volume} {33}},\ \bibinfo {pages} {213001} (\bibinfo {year}
  {2016}{\natexlab{a}})}\BibitemShut {NoStop}%
\bibitem [{\citenamefont {Celada}\ \emph
  {et~al.}(2016{\natexlab{b}})\citenamefont {Celada}, \citenamefont
  {Gonz\'alez},\ and\ \citenamefont {Montesinos}}]{cgm-PlebLike-2016}%
  \BibitemOpen
  \bibfield  {author} {\bibinfo {author} {\bibfnamefont {M.}~\bibnamefont
  {Celada}}, \bibinfo {author} {\bibfnamefont {D.}~\bibnamefont {Gonz\'alez}},
  \ and\ \bibinfo {author} {\bibfnamefont {M.}~\bibnamefont {Montesinos}},\
  }\href {\doibase 10.1103/PhysRevD.93.104058} {\bibfield  {journal} {\bibinfo
  {journal} {Phys. Rev. D}\ }\textbf {\bibinfo {volume}
  {\href{http://link.aps.org/doi/10.1103/PhysRevD.93.104058}{93}}},\ \bibinfo
  {pages} {104058} (\bibinfo {year} {2016}{\natexlab{b}})}\BibitemShut
  {NoStop}%
\bibitem [{\citenamefont {Herfray}\ and\ \citenamefont
  {Krasnov}()}]{Krasn2015}%
  \BibitemOpen
  \bibfield  {author} {\bibinfo {author} {\bibfnamefont {Y.}~\bibnamefont
  {Herfray}}\ and\ \bibinfo {author} {\bibfnamefont {K.}~\bibnamefont
  {Krasnov}},\ }\href@noop {} {}\Eprint {http://arxiv.org/abs/1503.08640}
  {arXiv:1503.08640} \BibitemShut {NoStop}%
\bibitem [{\citenamefont {Ashtekar}(1986)}]{Ashprl57.2244}%
  \BibitemOpen
  \bibfield  {author} {\bibinfo {author} {\bibfnamefont {A.}~\bibnamefont
  {Ashtekar}},\ }\href {\doibase 10.1103/PhysRevLett.57.2244} {\bibfield
  {journal} {\bibinfo  {journal} {Phys. Rev. Lett.}\ }\textbf {\bibinfo
  {volume} {57}},\ \bibinfo {pages} {2244} (\bibinfo {year}
  {1986})}\BibitemShut {NoStop}%
\bibitem [{\citenamefont {Ashtekar}(1987)}]{AshPRD.36.1587}%
  \BibitemOpen
  \bibfield  {author} {\bibinfo {author} {\bibfnamefont {A.}~\bibnamefont
  {Ashtekar}},\ }\href {\doibase 10.1103/PhysRevD.36.1587} {\bibfield
  {journal} {\bibinfo  {journal} {Phys. Rev. D}\ }\textbf {\bibinfo {volume}
  {36}},\ \bibinfo {pages} {1587} (\bibinfo {year} {1987})}\BibitemShut
  {NoStop}%
\bibitem [{\citenamefont {Capovilla}\ \emph {et~al.}(1991)\citenamefont
  {Capovilla}, \citenamefont {Dell}, \citenamefont {Jacobson},\ and\
  \citenamefont {Mason}}]{capo1991841}%
  \BibitemOpen
  \bibfield  {author} {\bibinfo {author} {\bibfnamefont {R.}~\bibnamefont
  {Capovilla}}, \bibinfo {author} {\bibfnamefont {J.}~\bibnamefont {Dell}},
  \bibinfo {author} {\bibfnamefont {T.}~\bibnamefont {Jacobson}}, \ and\
  \bibinfo {author} {\bibfnamefont {L.}~\bibnamefont {Mason}},\ }\href
  {http://stacks.iop.org/0264-9381/8/i=1/a=009} {\bibfield  {journal} {\bibinfo
   {journal} {Classical Quantum Gravity}\ }\textbf {\bibinfo {volume} {8}},\
  \bibinfo {pages} {41} (\bibinfo {year} {1991})}\BibitemShut {NoStop}%
\bibitem [{\citenamefont {Peld\'an}(1994)}]{peld1994115}%
  \BibitemOpen
  \bibfield  {author} {\bibinfo {author} {\bibfnamefont {P.}~\bibnamefont
  {Peld\'an}},\ }\href {http://stacks.iop.org/0264-9381/11/i=5/a=003}
  {\bibfield  {journal} {\bibinfo  {journal} {Classical Quantum Gravity}\
  }\textbf {\bibinfo {volume} {11}},\ \bibinfo {pages} {1087} (\bibinfo {year}
  {1994})}\BibitemShut {NoStop}%
\bibitem [{\citenamefont {Urbantke}(1984)}]{urba198425}%
  \BibitemOpen
  \bibfield  {author} {\bibinfo {author} {\bibfnamefont {H.}~\bibnamefont
  {Urbantke}},\ }\href
  {http://scitation.aip.org/content/aip/journal/jmp/25/7/10.1063/1.526402}
  {\bibfield  {journal} {\bibinfo  {journal} {J. Math. Phys.}\ }\textbf
  {\bibinfo {volume} {25}},\ \bibinfo {pages} {2321} (\bibinfo {year}
  {1984})}\BibitemShut {NoStop}%
\bibitem [{\citenamefont {Capovilla}\ \emph {et~al.}(1990)\citenamefont
  {Capovilla}, \citenamefont {Jacobson},\ and\ \citenamefont
  {Dell}}]{Capo-7-1-001}%
  \BibitemOpen
  \bibfield  {author} {\bibinfo {author} {\bibfnamefont {R.}~\bibnamefont
  {Capovilla}}, \bibinfo {author} {\bibfnamefont {T.}~\bibnamefont {Jacobson}},
  \ and\ \bibinfo {author} {\bibfnamefont {J.}~\bibnamefont {Dell}},\ }\href
  {http://stacks.iop.org/0264-9381/7/i=1/a=001} {\bibfield  {journal} {\bibinfo
   {journal} {Classical Quantum Gravity}\ }\textbf {\bibinfo {volume} {7}},\
  \bibinfo {pages} {L1} (\bibinfo {year} {1990})}\BibitemShut {NoStop}%
\bibitem [{\citenamefont {Torre}(1990)}]{Torreprd41}%
  \BibitemOpen
  \bibfield  {author} {\bibinfo {author} {\bibfnamefont {C.~G.}\ \bibnamefont
  {Torre}},\ }\href {\doibase 10.1103/PhysRevD.41.3620} {\bibfield  {journal}
  {\bibinfo  {journal} {Phys. Rev. D}\ }\textbf {\bibinfo {volume} {41}},\
  \bibinfo {pages} {3620} (\bibinfo {year} {1990})}\BibitemShut {NoStop}%
\bibitem [{\citenamefont {Samuel}(1988)}]{Samuel1988}%
  \BibitemOpen
  \bibfield  {author} {\bibinfo {author} {\bibfnamefont {J.}~\bibnamefont
  {Samuel}},\ }\href {http://stacks.iop.org/0264-9381/5/i=8/a=002} {\bibfield
  {journal} {\bibinfo  {journal} {Classical Quantum Gravity}\ }\textbf
  {\bibinfo {volume} {5}},\ \bibinfo {pages} {L123} (\bibinfo {year}
  {1988})}\BibitemShut {NoStop}%
\bibitem [{\citenamefont {Hojman}\ \emph {et~al.}(1976)\citenamefont {Hojman},
  \citenamefont {Kucha\v{r}},\ and\ \citenamefont {Teitelboim}}]{Hojman}%
  \BibitemOpen
  \bibfield  {author} {\bibinfo {author} {\bibfnamefont {S.~A.}\ \bibnamefont
  {Hojman}}, \bibinfo {author} {\bibfnamefont {K.}~\bibnamefont {Kucha\v{r}}},
  \ and\ \bibinfo {author} {\bibfnamefont {C.}~\bibnamefont {Teitelboim}},\
  }\href {\doibase http://dx.doi.org/10.1016/0003-4916(76)90112-3} {\bibfield
  {journal} {\bibinfo  {journal} {Ann. Phys.}\ }\textbf {\bibinfo {volume}
  {96}},\ \bibinfo {pages} {88} (\bibinfo {year} {1976})}\BibitemShut {NoStop}%
\bibitem [{\citenamefont {Bengtsson}(1991{\natexlab{a}})}]{beng19913158}%
  \BibitemOpen
  \bibfield  {author} {\bibinfo {author} {\bibfnamefont {I.}~\bibnamefont
  {Bengtsson}},\ }\href {\doibase 10.1063/1.529473} {\bibfield  {journal}
  {\bibinfo  {journal} {J. Math. Phys.}\ }\textbf {\bibinfo {volume} {32}},\
  \bibinfo {pages} {3158} (\bibinfo {year} {1991}{\natexlab{a}})}\BibitemShut
  {NoStop}%
\bibitem [{\citenamefont {Capovilla}\ \emph {et~al.}(1989)\citenamefont
  {Capovilla}, \citenamefont {Jacobson},\ and\ \citenamefont
  {Dell}}]{capo198963}%
  \BibitemOpen
  \bibfield  {author} {\bibinfo {author} {\bibfnamefont {R.}~\bibnamefont
  {Capovilla}}, \bibinfo {author} {\bibfnamefont {T.}~\bibnamefont {Jacobson}},
  \ and\ \bibinfo {author} {\bibfnamefont {J.}~\bibnamefont {Dell}},\ }\href
  {\doibase 10.1103/PhysRevLett.63.2325} {\bibfield  {journal} {\bibinfo
  {journal} {Phys. Rev. Lett.}\ }\textbf {\bibinfo {volume} {63}},\ \bibinfo
  {pages} {2325} (\bibinfo {year} {1989})}\BibitemShut {NoStop}%
\bibitem [{\citenamefont {Bengtsson}(1991{\natexlab{b}})}]{beng1991254}%
  \BibitemOpen
  \bibfield  {author} {\bibinfo {author} {\bibfnamefont {I.}~\bibnamefont
  {Bengtsson}},\ }\href {\doibase 10.1016/0370-2693(91)90395-7} {\bibfield
  {journal} {\bibinfo  {journal} {Phys. Lett. B}\ }\textbf {\bibinfo {volume}
  {254}},\ \bibinfo {pages} {55 } (\bibinfo {year}
  {1991}{\natexlab{b}})}\BibitemShut {NoStop}%
\bibitem [{\citenamefont {Krasnov}(2008)}]{KransDef}%
  \BibitemOpen
  \bibfield  {author} {\bibinfo {author} {\bibfnamefont {K.}~\bibnamefont
  {Krasnov}},\ }\href {\doibase 10.1103/PhysRevLett.100.081102} {\bibfield
  {journal} {\bibinfo  {journal} {Phys. Rev. Lett.}\ }\textbf {\bibinfo
  {volume} {100}},\ \bibinfo {pages} {081102} (\bibinfo {year}
  {2008})}\BibitemShut {NoStop}%
\bibitem [{\citenamefont {Krasnov}(2011{\natexlab{c}})}]{PhysRevD.84.024034}%
  \BibitemOpen
  \bibfield  {author} {\bibinfo {author} {\bibfnamefont {K.}~\bibnamefont
  {Krasnov}},\ }\href {\doibase 10.1103/PhysRevD.84.024034} {\bibfield
  {journal} {\bibinfo  {journal} {Phys. Rev. D}\ }\textbf {\bibinfo {volume}
  {84}},\ \bibinfo {pages} {024034} (\bibinfo {year}
  {2011}{\natexlab{c}})}\BibitemShut {NoStop}%
\end{thebibliography}%

\end{document}